\documentclass[floatfix,aps,prb,twocolumn,showpacs]{revtex4-1}
\usepackage{graphicx}
\usepackage[]{epsfig}
\usepackage{times,amsmath,amssymb}
\begin{document}
\title{Evidence of Spin-Filtering in Quantum Constrictions with Spin-Orbit Interaction}

\author{Sunwoo Kim}
\email{swkim@issp.u-tokyo.ac.jp}
\author{Yoshiaki Hashimoto}%
\author{Yasuhiro Iye}%
\author{Shingo Katsumoto}%

\affiliation{Institute for Solid State Physics, University of Tokyo, 5-1-5 Kashiwanoha, Kashiwa, Chiba 277-8581, Japan}

\date{\today}
\begin{abstract}
A new type of blockade effect $-$ spin-orbit blockade (SOB) $-$ has been observed in the conduction of a quantum dot (QD) made of a material with spin-orbit interaction. The blockade arises from spin-filtering effect in a quantum point contact (QPC), which is a component of the QD. Hence the appearance of the blockade itself evidences the spin-filtering effect in the QPC. The lower bound of filtering efficiency is estimated to be above 80\%.
\end{abstract}

\maketitle

Generation of spin current without external magnetic field in non-magnetic materials is a most important issue in semiconductor spintronics.~\cite{2002AwschSprin} Combination of spin-orbit interaction (SOI) and quantum constrictions is a promising candidate for spin-filtering and numbers of device schemes have been proposed~\cite{2005YamaPRB,2005EtoJPSJ,2008AhaPRB} though no clear experimental support has been given to them. A major difficulty lies in the detection of spin polarization through such delicate filters. In this article, we present the first clear observation of spin filtering with a simple quantum point contact (QPC) made of a material with SOI~\cite{2009DebNatNano} through finding of a novel blockade effect in a quantum dot (QD). We here call the phenomenon spin-orbit blockade (SOB). SOB is similar to spin blockade, in that it originates from the Pauli exclusion principle~\cite{2002OnoSci} though it requires only a single QD while ordinal spin blockade needs two QDs in series. Furthermore SOB gives estimation of the lower bound on the degree of polarization. We thus show that a QPC and a QD with SOI can work as a spin polarizer and a detector respectively.

SOI in a two-dimensional electron system (2DES) works on spins as an effective magnetic field, which arises from motion of electrons in electric fields. Broken spatial inversion symmetry, {\it e.g}., lattice asymmetry (Dresselhouse SOI) or asymmetric interfaces (Rashba SOI)~\cite{2003WinkSprin} can be the origin of built-in electric field. SOI is comparatively stronger in materials with smaller band gaps such as InAs, InSb, and mixed crystal (In,Ga)As. They are often chosen as stages for the investigation of SOI in semiconductor quantum structures.

Here we adopt a pseudomorphic In${}_{0.1}$Ga${}_{0.9}$As quantum well placed next to a GaAs/AlGaAs interface, which produces a perpendicular electric field. The layered structure shown in Fig.1(a) was grown with molecular beam epitaxy on a (001) GaAs substrate. The 2DES carrier concentration is 1.2$\times$10${}^{12}$/cm${}^{2}$, and the Hall mobility is 8.2$\times$10${}^{4}$cm${}^{2}$/Vs at 4.2K. QPCs and QDs defined by Ti/Au split gates were fabricated with electron-beam lithography. The specimens were cooled down to 50mK in a dilution fridge.

Figure 1(b) shows the conductance of a single QPC (gate configuration in Fig.1(a)) as a function of the split gate voltage. A clear plateau at half conductance quantum (0.5\textit{G}${}_{\rm q}$, \textit{G}${}_{\rm q}$=2\textit{e}${}^{2}$/\textit{h}) (0.5 plateau) is observed and the result is similar to that presented by Debray et al~\cite{2009DebNatNano} on a QPC also with strong SOI. They explained the 0.5 plateau along the interaction-driven spin-filtering though no further experimental support has been given other than the response to magnetic field and to asymmetry in the QPC potential.
\begin{figure}
\begin{center}
\includegraphics{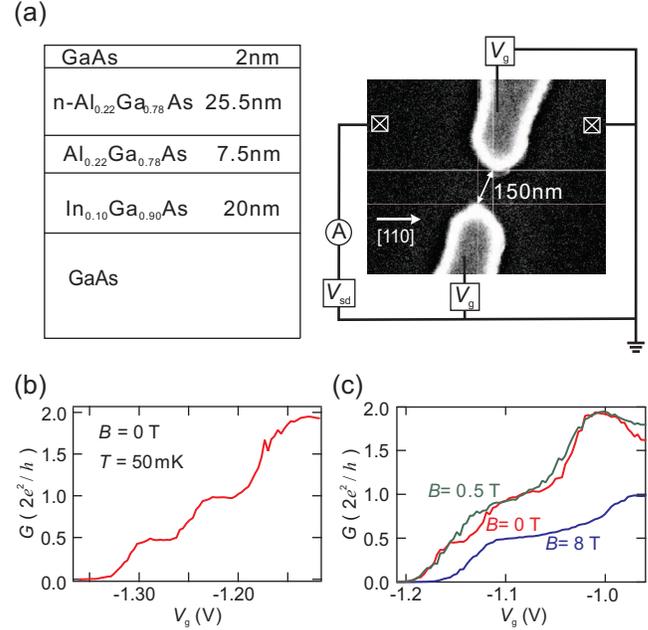}
\caption{(Color online) (a) Schematic cross-sectional view of layered structure grown with MBE. Nominal Si concentration in the modulation doped layer is 6$\times$10${}^{18}$$/$cm${}^{3}$. The right figure is the electron beam micrograph of metallic gate configuration for QPC. (b) Two wire conductance of a QPC at 50mK as a function of symmetrically applied split gate voltage. Separately measured contact resistance is subtracted. A clear 0.5\textit{G}${}_{\rm q}$ quantized plateau is observed. (c) Perpendicular magnetic field variation of the 0.5\textit{G}${}_{\rm q}$ plateau, which disappears at 0.5T and reappears as a broad plateau around 8T.}
\label{Fig1}
\end{center}
\end{figure}

Henceforth we give strong experimental supports to the hypothesis of spin-filtering. As a first simple check, the conduction under the perpendicular magnetic field up to 10T was examined. Above 7T the Zeeman splitting is large enough to dominate the subband structure at the QPC and a clear 0.5\textit{G}${}_{\rm q}$ plateau appears as shown in Fig.1(c) confirming that the spin-filtering can cause half-quantum discretization. Interestingly the 0.5\textit{G}${}_{\rm q}$ plateau once disappears around 0.5T. According to Ref.5, asymmetric confinement potential is indispensable for spin-filtering and if similar is applicable to the present case, a weak perpendicular field would reduce boundary scattering~\cite{1991AkePRB} and resultant ``recovery'' of symmetry is a possible cause of the disappearance. 
\begin{figure}
\begin{center}
\includegraphics{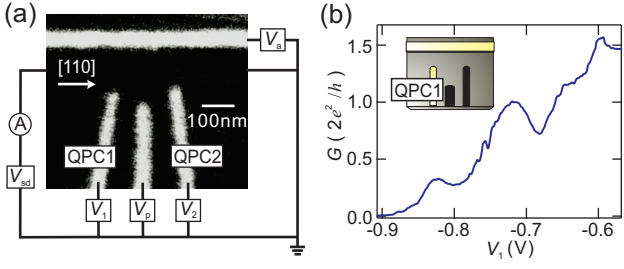}
\caption{(Color online) (a) Scanning electron micrograph of split gate configuration for QD. Measurement circuit is schematically drawn.
(b) Conductance of a QPC1 at 50mK as a function of \textit{V}${}_{\rm 1}$ with \textit{V}${}_{\rm a}$=$-$0.7V and \textit{V}${}_{\rm p}$,\textit{ V${}_{\rm 2}$}=0V.}
\label{Fig2}
\end{center}
\end{figure}

Though the above is in agreement with the hypothesis of spin-filtering in the QPC, support from more spin-specific phenomenon is required in order for proving that 0.5 plateau is really due to spin-filtering.  We hence proceed to the conduction through a QD with the same material. Through the clarification of electronic states in the QD, it can be possible to exclude possible physical origins other than spin-filtering. As shown in Fig.2(a), our QD is connected to the electrodes through two QPCs (QPC1, QPC2) on both sides. As depicted in Fig.2(b), these QPCs also show 0.5 plateau in the individual measurements. In forming a QD, QPC1 was set on 0.5 plateau while QPC2 was on 1.0 plateau. Therefore if we apply the spin-filtering hypothesis, the transport through QPC1 is spin-selective while that through QPC2 is not.

As noted above we should know the spin state in the dot for the examination of the hypothesis. Here we utilize the Kondo effect to get concrete information on the spin states.~\cite{2001KouwenPW} The Kondo effect in conduction through QDs is characterized by anomalous enhancement of conductance in Coulomb valleys with decreasing temperature and by zero-bias conductance peaks.~\cite{2000VanderSci} Because these features are the results of interaction between electron spins via localized spins in the dot, the QD should have an unpaired electron at the topmost level in the valleys with the Kondo effect. We looked for such a Coulomb valley among few tens of Coulomb oscillation periods.

In Fig.3(a), where we index Coulomb valleys from A to D, peaks from p${}_{\rm a}$ to p${}_{\rm d}$, the candidate for a Coulomb-Kondo valley is C. Firstly, Coulomb peaks at the both sides of C (p${}_{\rm c}$ and p${}_{\rm d}$) have similar and outstanding heights indicating that these two are through the same orbital states (\textit{i.e.}, Kramers degenerated), which have large coupling constant $\Gamma$ to the electrodes.~\cite{2000SilvePRL,2004AikaJPSJ} This means the Kondo temperature \textit{T}${}_{\rm K}$, which can be approximated in the mid-point of the valley with Coulomb repulsion parameter \textit{U} as 
\begin{equation} \label{GrindEQ__1_} 
T_{{\rm K}} =\frac{\sqrt{\Gamma U} }{2} \exp \left[-\frac{3\pi U}{4\Gamma } \right],         
\end{equation} 
\begin{figure}
\begin{center}
\includegraphics{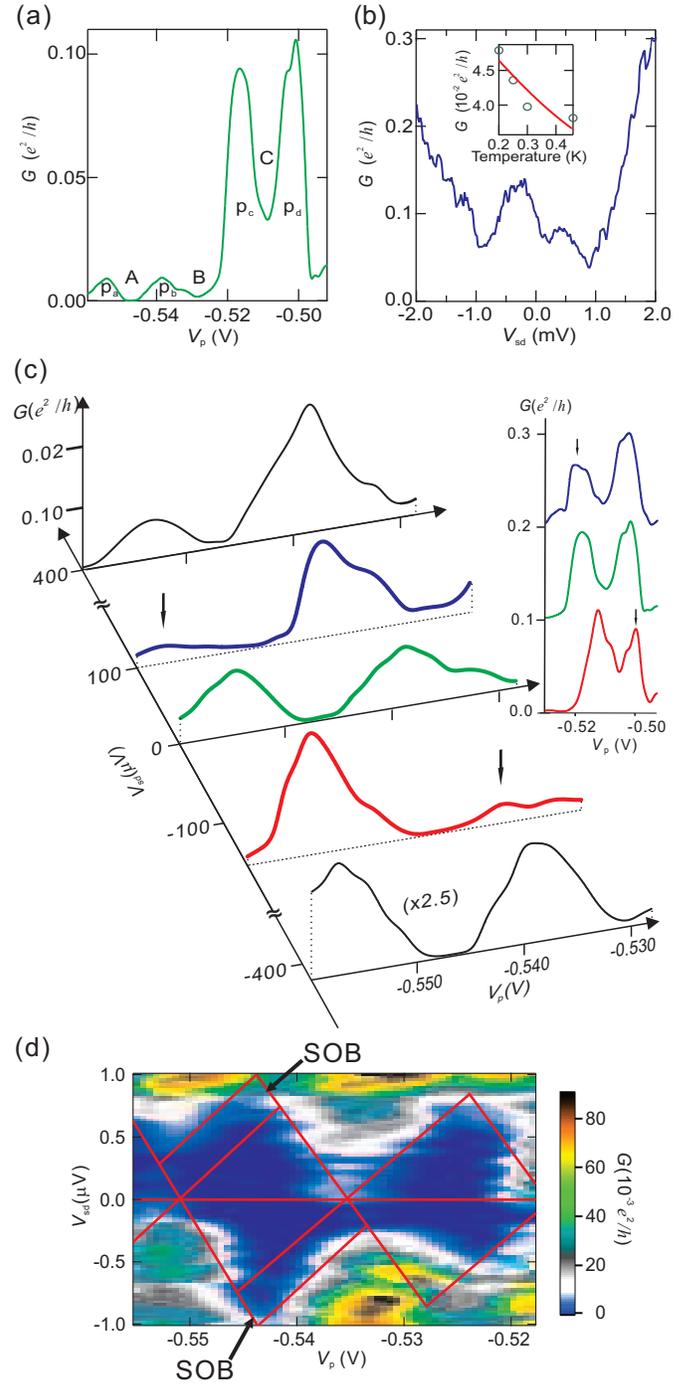}
\caption{(Color online)
(a) QD conductance at 50mK as a function of plunger gate voltage \textit{V}${}_{\rm p}$. Four Coulomb peaks are observed. The peaks and the valleys are labeled p${}_{\rm a}$-p${}_{\rm d}$, A-C respectively.
 (b) QD conductance at 50mK as a function of \textit{V}${}_{\rm sd}$ at the mid-point of Coulomb valley C. The inset shows temperature dependence of finite bias QD conductance and the line is the fitted result of Kondo temperature dependence with \textit{T}${}_{\rm K}$ of 995mK.
(c) Quantum dot conductance at 50mK as a function of plunger gate voltage \textit{V}${}_{\rm p}$${}_{ }$for five representative \textit{V}${}_{\rm sd}$ conditions. At \textit{V}${}_{\rm sd}$=0 (green curve) there are four Coulomb peaks in this range of \textit{V}${}_{\rm p}$ as indicated in the figure. At \textit{V}${}_{\rm sd}$ =100$\mu$V (blue curve), we observe peak p${}_{\rm a}$ disappears and p${}_{\rm c}$ diminishes as indicated by vertical arrows while at　\textit{V}${}_{\rm sd}$=$-$160$\mu$V(red curve) peak p${}_{\rm b}$ disappears and p${}_{\rm d}$ diminishes. Further increment of \textbar \textit{V}${}_{\rm sd}$\textbar  recovers the double peak configuration. The inset shows similar tendency in peaks p${}_{\rm c}$ and p${}_{\rm d}$.
(d) Color plot of the QD conductance on \textit{V}${}_{\rm p}$-\textit{V}${}_{\rm sd}$ plane around valleys A and C. The Coulomb diamonds are marked by black arrows. The SOB regions (see Fig.4(e)) are also indicated.}
\label{Fig3}
\end{center}
\end{figure}
is comparatively higher in C. Secondly the current-voltage (I-V) characteristic shows a broad double-peak structure at 50mK as displayed in Fig.3(b). This is similar to the I-V curve for the Kondo effect other than the zero-bias dip. As discussed later, the dip at the center is probably caused by the spin- filtering at QPC1. Lastly enhancement of the conductance with cooling is observed as shown in the inset of Fig.3(b). If we fit well known temperature dependence of conductance~\cite{1994CostiJPCM,1998GoldPRL} \textit{G}(\textit{T})=\textit{G}${}_{\rm 0}$$-$\textit{G}${}_{\rm 1}$[\textit{T}${}_{\rm K}$${}^{2}$/(15\textit{T}${}^{2}$+\textit{T}${}_{\rm K}$${}^{2}$)]${}^{1/4}$ for spin$-$1/2 Kondo effect to the data, we obtain reasonable \textit{T}${}_{\rm K}$ of about 1K. From the above analysis, we can safely say that the SU(2) (spin 1/2) Kondo effect is emerging at valley C.

With the knowledge of the spin state in valley C, the topmost state in valley A is identified as the one with an unpaired electron, while the dot must have a closed shell structure in valley B. This assignment does not change even if we take into account the possibility of a high spin state due to the electron correlation in the dot.
\begin{figure}
\begin{center}
\includegraphics{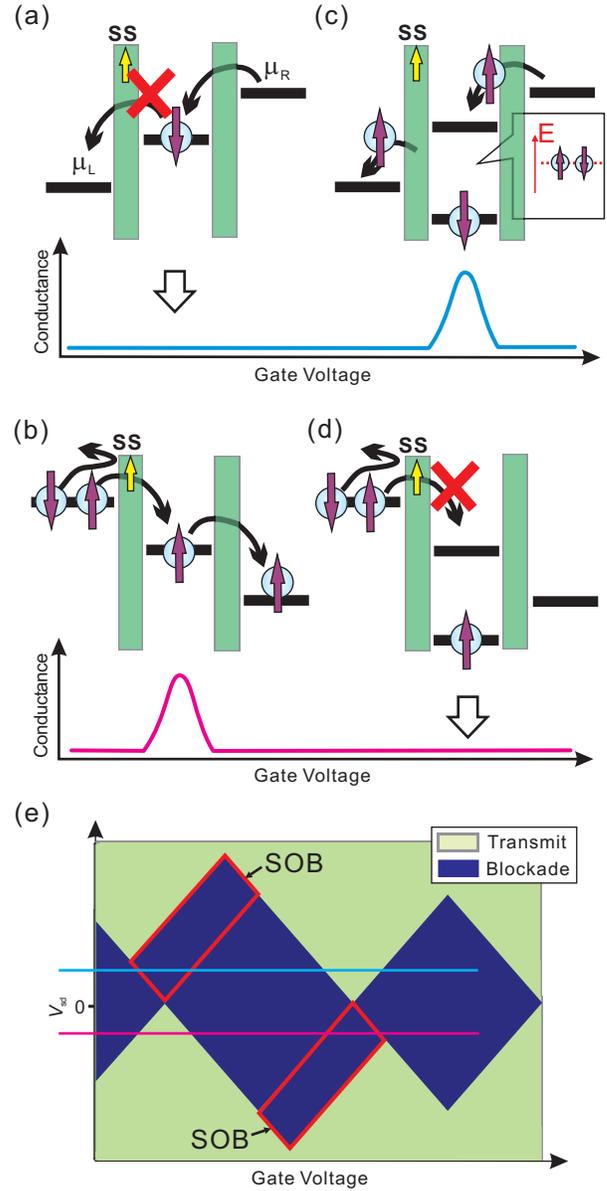}
\caption{(Color online)
(a) At peak p${}_{\rm a}$, when a positive bias is applied, the topmost level is empty before the supply of an electron from the drain through QPC2. Sooner or later the level is occupied by a $\downarrow$-electron, which cannot go out to the source due to the spin-selectivity in QPC1, and the current is blocked.
(b) When a negative bias is applied at p${}_{\rm a}$, the electron at the topmost level can go out to the drain through QPC2 regardless of the spin direction and the current flows.
(c) At peak p${}_{\rm b}$, when a positive bias voltage is applied, the topmost level is already occupied with an electron before the supply of another electron with opposite direction from the drain through QPC2. After the entrance the level is occupied with a singlet pair and the $\uparrow$-electron can go out to the source through QPC1 with spin selectivity.
(d) When a negative bias is applied at p${}_{\rm b}$, QPC1 can supply only $\uparrow$-electrons and sooner or later a $\downarrow$-electron goes out to the drain and successive tunneling is blocked by the Pauli exclusion principle.
(e) Summary of the above discussion to modified Coulomb diamond characteristics.}
\label{Fig4}
\end{center}
\end{figure}

Our target of analysis hereafter is valley A, where the bordering peaks (p${}_{\rm a}$ and p${}_{\rm b}$) are significantly lower than those of valley C, \textit{i.e}., $\Gamma$ is much smaller, hence from Eq.\eqref{GrindEQ__1_} \textit{T}${}_{\rm K}$ is very low. The Kondo effect is thus out of experimental scope here. We observe normal Coulomb oscillation (green curve in Fig.3(c)) at zero-bias (\textit{V}${}_{\rm sd}$=0). At a small positive bias (\textit{V}${}_{\rm sd}$=100$\mu$V) where broadening and height growing of the Coulomb peaks are expected for increasing \textbar \textit{V}${}_{\rm sd}$\textbar , surprisingly peak p${}_{\rm a}$ vanishes while p${}_{\rm b}$ shows ordinal growth (blue curve in Fig.3(c)). This is reminiscent of spin-blockade in double QD systems.~\cite{2002OnoSci} However in the present case, when the sign of \textit{V}${}_{\rm sd}$ is reversed to negative, peak p${}_{\rm b}$ vanishes whereas p${}_{\rm a}$ grows as usual (red curve in Fig.3(c)). These anomalous behaviors disappear and ordinal Coulomb oscillation is restored when the absolute value of \textbar \textit{V}${}_{\rm sd}$\textbar  exceeds about 250$\mu$V (black curves in Fig.3(c)). The results are summarized in Fig.3(d).

Because p${}_{\rm a}$ and p${}_{\rm b}$ arise from the same orbital level, the major difference between them is the total spin in the dot. In the following we discuss that the spin-selectivity (SS) at QPC1 is the origin of the unexpected anti-symmetric blockade. For simplicity we henceforth assume that only up spin ($\uparrow$) electron can transmit through QPC1 without losing generality. The following discussion also holds if an entanglement between the direction of the electron transmission and the spin (\textit{e.g.}, left-$\uparrow$ and right-$\downarrow$) is assumed instead of the simple SS.
Therefore we cannot go into the comparison of the present results with existing
theoretical proposals\cite{2009WanPRB,2010EntinPRB} for such spin-selectivity in quantum constrictions with SOI.

Let us begin with peak p${}_{\rm b}$, where the topmost orbital level is already occupied by an unpaired electron. When a positive \textit{V}${}_{\rm sd}$ is applied, the flow sequence of electric current through the QD is as follows: first an electron tunnels from the drain into the QD through QPC2, then only $\uparrow$-electron can escape through QPC1 with $\uparrow$-SS but another $\uparrow$-electron can be supplied through QPC2 (Fig.4(c)). Hence the process forms a finite Coulomb peak. When the bias is reversed, electrons should come in through QPC1 and go out through QPC2. When QPC2 transmits a $\downarrow$-electron, the topmost level is left with an $\uparrow$-electron and the following entrance of $\uparrow$-electron through QPC1 is blocked due to the Pauli exclusion principle. The current is hence blocked by the $\uparrow$-SS of QPC1 (Fig.4(d)). Spin-flips in the dot and leakage of $\uparrow$-SS of QPC1 appear as a leak current.

At peak p${}_{\rm a}$ for positive \textit{V}${}_{\rm sd}$, the flow sequence is: QPC2 $\rightarrow$ the lowest empty state above the closed shell $\rightarrow$ QPC1. When a $\downarrow$-electron comes into the dot it cannot escape due to the $\uparrow$-SS of QPC1 and the flow is blocked (Fig.4(a)). The blockade does not take place for negative \textit{V}${}_{\rm sd}$ because $\uparrow$-electrons supplied through QPC1 can escape through QPC2 (Fig.4(b)). This anti-symmetric blockade, what we call SOB here, deduced from the simple assumption of SS in QPC1 perfectly agrees with the observations in Fig.3.

The SOB can be lifted with further increase in the amplitude of \textit{V}${}_{\rm sd}$. With increasing \textbar \textit{eV}${}_{\rm sd}$\textbar , the energy gain from the voltage source associated with tunneling increases and when it exceeds the spacing of single-electron orbital levels, multi-level transport begins forming spin-bypaths; or when it exceeds the gap between $\uparrow$ and $\downarrow$ channels, QPC1 is transparent for both spins. Therefore the SOB is lifted along a line parallel to an edge of a Coulomb diamond resulting in anti-symmetric Coulomb diamonds as illustrated in Fig.4(e).

Note that the same SOB mechanism is alive even in the Kondo region, though it is not so prevailing due to large $\Gamma$ (broad level) and frequent spin fluctuation due to the Kondo effect. If we focus on the Coulomb peak heights, the same tendency to those of peaks p${}_{\rm a}$ and p${}_{\rm b}$ is observed for p${}_{\rm c}$ and p${}_{\rm d}$ (blue and red curves in Fig.3(c)). Peak-splitting like outlook of the Kondo peak in the I-V characteristics in Fig.3(b) can be interpreted along the same line. At zero-bias, the Kondo enhancement of co-tunneling is only available for QPC2 because \textit{T}${}_{\rm K}$ for co-tunneling over QPC1 is reduced, in the sense of Eq.\eqref{GrindEQ__1_}, by exp($-\pi$\textit{e}${}^{2}$\textit{V}${}_{\rm th}$${}^{2}$/$\Gamma$\textit{U}), where \textit{V}${}_{\rm th}$ is the SOB lifting threshold in \textit{V}${}_{\rm sd}$. With increasing \textbar \textit{V}${}_{\rm sd}$\textbar , this reduction is suppressed and the ordinal conductance enhancement of the Kondo effect is restored. As a result, a broad double peak structure in Fig.3(b) appears. All the above deductions are summarized in Fig.4(e) and we find every characteristic point is in good agreement with Fig.3(a)-(d).

So far we have proven that spin-filtering is realized in QPCs with SOI on 0.5 plateaus and that can be detected through the appearance of SOB in the conduction of QDs. It should be noted that the peak height ``switching'' ratios in SOB peaks (in the present case p${}_{\rm a}$ and p${}_{\rm b}$) give the lower bound of spin-filtering efficiency in QPC1, which should be above 80\% in Fig.3. It is surprising that such high efficiency is obtained even with such a small In content (10\%). A possible reason is the high 2DES concentration, which makes the gap between opposite spins at the Fermi circle (\textit{i.e.}, for fixed momentum) very large. Actually we do not observe quantization in 2DES of the same material but with much lower concentration. The present scheme for spin-filtered polarization detection is widely applicable even apart from SOI, to various problems (\textit{e.g}., to the 0.7 problem).

In summary, we have confirmed conductance quantization by half-conductance quantum in QPCs made of In${}_{0.1}$Ga${}_{0.9}$As 2DES. We have found a QD, which consists of two such QPC, shows characteristic conductance blockade (SOB), which arises from spin-selective tunneling due to the SOI. The first observation of the SOB directly evidences very high spin-filtering efficiency at QPCs with SOI. QPCs with SOI can work as, in other words, spin polarizers and QDs as detectors.
\newline

This work is supported by Grant-in-Aid for Scientific Research and Special Coordination Funds for Promoting Science and Technology.

\end{document}